\documentclass[useAMS,usenatbib]{mn2e}
\usepackage{graphicx}
\usepackage{epsf}
\newcommand{\Mo}{M_{\odot }}
\newcommand{\apj}{ApJ}
\newcommand{\apjl}{ApJL}
\newcommand{\apjs}{ApJS}
\newcommand{\mnras}{MNRAS}
\newcommand{\aj}{AJ}
\newcommand{\araa}{ARA\&A}
\newcommand{\nat}{Nature}
\newcommand{\aap}{A\&A}
\newcommand{\aaps}{A\&A Supplement}

\title[The dark GRB080207]
{The dark GRB080207 in an extremely red host and the implications for GRBs in highly obscured environments}

\author[K.M.~Svensson et al.]{K.M.~Svensson$^{1}$,
A.J.Levan$^{1}$\thanks{email: a.j.levan@warwick.ac.uk}, N.R. Tanvir$^{2}$, D.A. Perley$^{3}$, M.J. Michalowski$^{4}$, \and  K.L. Page$^{2}$,
J.S. Bloom$^{3}$, S.B. Cenko$^{3}$, J. Hjorth$^{5}$, P. Jakobsson$^{6}$,  \and D. Watson$^{5}$, P.J. Wheatley$^{1}$ \\
$^1$Department of Physics, University of Warwick, Coventry, CV4 7AL, UK \\
$^{2}$Department of Physics and Astronomy, University of Leicester,
Leicester, LE1~7RH, UK. \\
$^{3}$Department of Astronomy, University of California, Berkeley, CA 94720-3411, USA\\
$^{4}$Scottish Universities Physics Alliance, Institute for Astronomy, University of Edinburgh, Royal Observatory, Edinburgh, EH9 EHJ, UK\\
$^{5}$Dark Cosmology Centre, Niels Bohr Institute, University of Copenhagen, Juliane Maries Vej 30, DK-2100 Copenhagen, Denmark \\
$^{6}$Centre for Astrophysics and Cosmology, Science Institute, University of Iceland, Dunhagi 5, IS-107, Iceland \\
}

\begin{document}

\date{Submitted 21st July 2011, in original form 21 October 2010.}

\pagerange{\pageref{firstpage}--\pageref{lastpage}} \pubyear{2002}

\maketitle

\label{firstpage}

\begin{abstract}
We present comprehensive X-ray, optical, near- and mid-infrared, and sub-mm observations of
GRB 080207 and its host galaxy. The afterglow was undetected in the optical and near-infrared (nIR) implying 
an X-ray-to-optical spectral slope less than 0.3, identifying GRB 080207 as a dark burst. 
{\em Swift} X-ray observations show extreme absorption in the host, which is confirmed by the unusually large
optical extinction found by modelling our the X-ray to nIR afterglow spectral energy distribution. 
Our {\em Chandra} observations obtained 8 days post-burst allow us to place the afterglow on the
sky to sub-arcsec accuracy, enabling us to pinpoint an extremely red galaxy (ERO), with R-K$>$5.4 (g-K $\sim 7.5$, Vegamag) at the afterglow
location. Follow-up host observations with the {\em Hubble Space Telescope, Spitzer Space Telescope},
Gemini, Keck and the James Clerk Maxwell Telescope (JCMT)
 provide a photometric redshift solution of $z\approx1.74^{+0.05}_{-0.06}$ ($1 \sigma$, $1.56 < z < 2.08$ at 2$\sigma$) for
the ERO host, and suggest that it is a massive and morphologically disturbed 
ultra-luminous infrared galaxy (ULIRG) system, with $L_{FIR} \sim 2.4 \times 10^{12}$ L$_{\odot}$. These results add to the growing 
evidence that GRBs originating in very red hosts always show some evidence of dust extinction in their afterglows
(though the converse is not true -- some extinguished afterglows are found in blue hosts). This indicates that a
poorly constrained fraction of GRBs occur in very dusty environments. By comparing the inferred stellar masses, and estimates of
the gas phase metallicity in both GRB hosts and sub-mm galaxies we suggest that many GRB hosts, even at $z>2$ are
at lower
metallicity than the sub-mm galaxy population, offering a likely explanation for the dearth of sub-mm detected GRB hosts. 
However, we also show that the dark GRB hosts are systematically more massive than those hosting optically bright events,
perhaps implying that previous host samples are severely biased by the exclusion of dark events. 
\end{abstract}


\section{Introduction}


A fraction of gamma-ray burst afterglows are undetected or have suppressed flux 
in the optical and
even in the nIR \cite[e.g.][]{1998ApJ...493L..27G}. These bursts may include 
high-redshift events 
or where there is significant absorption in the host galaxy. Alternatively, 
observational selection effects may result in a non-detection due to 
unfavourable location, poor weather etc. for ground based observatories.  These 
observational selection effects can largely be avoided by 
selecting bursts based on some quantitative criteria, in particular by 
comparing the optical limits on the afterglow emission to
the expected values based on the observed X-ray flux and spectral slope. By 
this criterion \cite{2004ApJ...617L..21J} \citep[see 
also][]{2005ApJ...624..868R} 
define dark bursts as those with an X-ray-to-optical spectral slope, 
$\beta_{OX}<0.5$, where $F_{\nu} \propto \nu^{-\beta}$ and 
\begin{equation}
        \beta_{OX}=\frac{\log_{10}(F_{\nu,X}/F_{\nu,Opt})}{\log_{10}(\nu_{X}/\nu_{Opt})}.
\end{equation}
In the range $0.5 < \beta_{OX} < 1.25$ which is suggested by the standard
fireball model, the distribution of $\beta_{OX}$ is approximately flat 
\citep[e.g. Figure 1 in ][]{2004ApJ...617L..21J}, with a tail of 
$\beta_{OX} < 0.5$ outliers.
\cite{2009ApJ...699.1087V} proposed a more sophisticated approach and define 
dark bursts
by  $\beta_{OX}<\beta_X-0.5$. Selecting bursts which are dark by these 
requirements, ensures the sample studied appears
genuinely physically distinct from the optically bright GRBs, in contrast to 
simple requirement of an optical non-detection, 
which is often not constraining in terms of physical models of the afterglow 
\citep{2005ApJ...624..868R}. 
Understanding these dark bursts, and the physical causes of darkness is 
important, not only for characterising the diversity of GRBs themselves, 
but also for establishing their utility as cosmological probes, and in 
particular as tracers of the global star formation rate. 

Since long GRBs are known to be associated with massive stars 
\citep[e.g.][]{2003ApJ...591L..17S,2003Natur.423..847H},
we can contemplate an ideal scenario
in which there was direct proportionality between GRB rate and star formation 
rate, allowing the GRB rate to be an immediate measure
of  the global star formation rate across cosmic history. Two particular advantages of 
GRBs in this role come from their brightness, allowing
them to be seen at the most extreme redshifts 
\citep{2009Natur.461.1254T,2009Natur.461.1258S}
and their high energy emission, enabling them to be seen through high dust 
columns. Coupled with this, they select galaxies across the
luminosity function (rather than just at the bright end). Hence, GRBs have the 
potential to infer the star formation rate, largely
free from the order of magnitude corrections that other techniques must apply.  
For example, estimates based on Lyman break galaxies must make corrections 
to account for missing galaxies at the faint end 
of the luminosity function \citep[e.g.]{2006ApJ...653...53B,2010MNRAS.409..855B}, and dust obscuration 
\citep[e.g.]{2009ApJ...705..936B,2011ApJ...726L...7O}. In contrast, sub-mm searches find the
most extreme, dusty examples, but (at least at present) cannot study fainter galaxies, 
and hence also require large scale corrections to obtain a total star formation rate 
\citep{1998Natur.394..241H,1999MNRAS.302..632B}. In practice however, the
promise of GRBs remains to be fulfilled. This is largely due to a combination
of incompleteness in the available samples 
\cite[e.g.][]{2006A&A...447..897J,2009ApJS..185..526F} 
for example because of the difficulty in locating dust obscured GRBs, and 
because of poorly known environmental effects 
\citep[such as metallicity, e.g.][]{2007MNRAS.375.1049W,2008AJ....135.1136M} 
on the GRB progenitors which impact any direct proportionality between GRB rate and star formation rate. 
An understanding of dark bursts offers a route through both of these problems; 
by increasing the completeness of GRB samples, the ability to obtain an 
accurate redshift distribution for the whole of the GRB population currently 
detected by {\em Swift} is gained. In tandem, studies of the environments of dark bursts, 
in comparison with those of bright examples can be extremely valuable in 
elucidating the impact of environment on GRB production. 

It is therefore reasonable to ask how studies of dark bursts can be achieved. 
GRBs are located in the gamma-rays and
subsequently pinpointed by their X-ray afterglows. Although X-ray afterglows 
in the \emph{Swift} era are ubiquitous, they
frequently do not allow detailed study of the burst due to the inability to 
obtain either absorption spectroscopy of the afterglow, or
the unambiguous detection of the host galaxy. Although {\em Swift} X-ray positioning has been greatly
improved by more refined algorithms that determine the satellites pointing using 
the Ultra-Violet and Optical Telescope (UVOT),
The median X-Ray Telescope (XRT) 
error circle is still $\sim 1.5$ arcsec, with 90\% of bursts being
positioned to less than $2$ arcsec \citep{2009MNRAS.397.1177E} - suggesting that the bulk of GRB host galaxies still can't be
unambiguously determined using only X-rays;
Purely by chance \cite[e.g. considering the galaxy number counts 
by][]{1997MNRAS.288..404H}, XRT error circles have $\sim 15\%$ probability of {\em 
randomly} containing a galaxy with $R<25$
-- roughly the median magnitude of GRB hosts \citep{1999ApJ...520...54H}, and 
may contain more than one galaxy comparable to the faintest known GRB 
hosts --  $R<29$ \citep{2006Natur.441..463F}. 
Hence, even the now well refined X-ray positions from the {\em Swift} X-ray 
Telescope \citep{2007A&A...469..379E} cannot unambiguously locate a host.
Although absorption in the X-ray afterglows can provide a clue to the GRB 
environment via the measurement 
of hydrogen column ($N_H$), this is one of few constraints that can be obtained 
from the X-ray
afterglow alone. Indeed, in the absence of a redshift, even the rest frame 
X-ray column cannot be accurately constrained. 
Although the definition of dark bursts doesn't require an optical afterglow 
non-detection, (and indeed in many cases the afterglow has been detected), 
selecting an unbiased sample of dark burst hosting galaxies calls for accurate 
identification of the host even in cases where the optical afterglow remains 
undetected. A possible solution to the problem of identifying the hosts is to 
obtain sub-arcsec astrometric positions, reducing the chance alignment by a factor of
$\sim 10$, for dark GRBs via their X-ray afterglows. Currently, this is only enabled 
by the {\em Chandra} X-ray Observatory, and this is the approach employed here. 

A consequence of the relative dearth of dark bursts (per the $\beta_{OX}<0.5$ 
definition rather than simply optical non-detection, also throughout this 
work) in the
pre-{\em Swift} era and relatively weak constraints which can be obtained from 
X-ray afterglows alone means that the
origins and hosts of dark GRBs remain relatively poorly understood, despite the 
relatively large number uncovered by {\em Swift}. It
is therefore not entirely clear how the environments (both local 
and galactic)  of dark GRBs differ from those
of the optically bright population. The fraction of dark bursts appears to be
$\sim 0.5$ \citep{2008ApJ...686.1209M,2009ApJ...693.1484C,2009ApJS..185..526F} 
with the majority of these being consistent with low to medium 
redshift events suffering from
dust extinction in the host \citep{2009AJ....138.1690P}, 
while perhaps $\sim 20\%$ originate from $z>5$ \citep{2011A&A...526A..30G}. 
This could significantly bias samples based on optical detections of the 
afterglow. 
Studying the host population of dark GRBs is therefore a priority in order to 
understand  
how they differ from normal bursts and what impact the difference will have on 
statistical host samples -- 
either by inclusion of dark burst hosts, or by their exclusion. Although the number of 
dark GRBs with securely identified hosts is still
relatively small it is noteworthy that several of other heavily extinguished bursts hitherto have 
been associated with galactic environments that are {\it atypical} of the
overall host population:  The hosts of GRB~020127 and GRB~030115 are massive
extremely red objects (EROs) \citep[e.g.][]{2007ApJ...660..504B,2006ApJ...647..471L}, although the $\beta_{OX}$ values are poorly or unconstrained due to
lacking follow-up observations \citep[][]{2002GCN..1250....1F,2005A&A...439..987S}. 
GRB~051022 has a massive host \citep{2010MNRAS.405...57S} with large average extinction
\citep[e.g.][]{2007ApJ...669.1098R,2007AA...475..101C}
and GRB~080325  also has a massive host with evidence of significant 
extinction \citep{2010arXiv1003.3717H}. 
Although this is not an exhaustive list of all dark bursts, in these cases the 
evidence seems to suggest
either unusually red hosts, unusually massive hosts or hosts with very high 
extinction. It is also interesting to note that, in the sample of 34 GRBs in 
\citep{2010MNRAS.405...57S} where stellar masses are estimated from spectral 
energy distribution (SED) fitting,
the hosts of all dark bursts are found to be above the median mass of the 
sample.

Here we present observations of GRB~080207 and its host, utilizing
multiwavelength observations stretching from the X-ray to the sub-mm to
identify a host galaxy, and study its properties in comparison to other 
dark, and bright, GRB hosts. 

\section{Observations}
GRB~080207 was discovered by {\em Swift} at 21:30:21 UT on 7 February 2008. A 
prompt slew enabled the location of an X-ray afterglow, however no optical 
afterglow was found in UVOT observations. The burst was long duration with 
$t_{90} > 300$s (at which point the source moved out of the BAT field of view 
\citep{2008GCN..7272....1S}).


\subsection{Afterglow}
\subsubsection{X-ray}
Observations with the {\em Swift} X-ray Telescope (XRT) began 124 seconds after 
the
burst, and continued for 30 hours. For spectral analysis the XRT 
observations were first processed through
{\tt xrtpipeline} to create cleaned event lists in both Window Timing (WT) 
and Photon
Counting (PC) mode. We separately fitted spectra to the WT and PC mode data using {\tt XSPEC}. The 
WT data are best fitted by an absorbed power-law model with spectral slope $\beta 
= 0.34 \pm 0.1$
($F_{\nu} \propto \nu^{-\beta}$), 
and $N_H = (96 \pm 11) \times 10^{20}$ cm$^{-2}$ (assuming zero redshift for 
the absorption), significantly in excess of 
the galactic value of  $1.94 \times 10^{20}$ cm$^{-2}$. The PC mode 
observations yielded a similar excess column density, $N_H = (75 \pm 16) \times 
10^{20}$ cm$^{-2}$ , but
a much softer spectral slope of $\beta = 1.4 \pm 0.1$. It is also worth noting that a consistently high $N_H$ for the
zero redshift case was also found by \cite{2008GCN..7266....1R}.

The WT mode observations took place during the period $130\textrm{ to }194$~s post 
burst. 
Throughout this time the BAT was also detecting higher energy emission, and the 
harder spectral index
measured in the WT data is most likely a consequence of the prompt emission 
in the
X-ray band. We therefore adopt the spectral slope of the afterglow as $\beta = 
1.4 \pm 0.1$, as measured in the PC mode observations. 

\begin{figure}
  \begin{center}
    \includegraphics[scale=0.4]{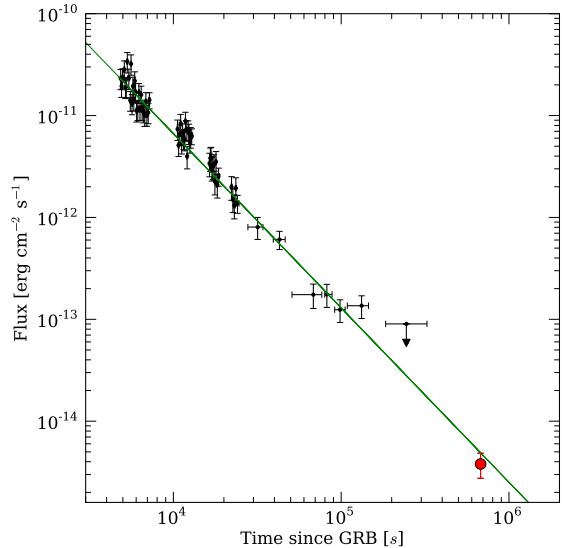}
    \caption[X-ray lightcurve of GRB~080207]{The X-ray lightcurve of 
      GRB~080207 from {\em Swift}/XRT PC mode (small black points) 
      and {\em Chandra} (large filled circle). 
      The {\it Chandra} flux  is rescaled from its observed ACIS bandwidth to 
      equivalent of the {\em Swift}/XRT in this figure. The solid green line shows 
      a single power law fit with a decay slope $\alpha = 1.7$.}
    \label{fig:lightcurve}
  \end{center}
\end{figure}

We took the X-ray lightcurve  from the {\em Swift} repository 
\citep{2007A&A...469..379E,2009MNRAS.397.1177E}, 
to which we added the {\it Chandra} observation at $t \sim 7 \times 10^5$~s. The 
lightcurve is roughly flat during the WT mode observations. The period between 
the end of WT and the beginning
of PC mode observations is broadly consistent with a single power law decay 
($F(t) \propto t^{-\alpha}$) of index $\alpha \sim 1.0$. There is no sign of, 
or requirement for steep
initial decays, or a later time plateau as seen in many X-ray afterglows 
\citep{2006ApJ...642..389N}. 
The PC mode late time (between 1000 and $10^6$~s)  
(Figure~\ref{fig:lightcurve}) is fitted with a single power law with a decay 
index
$\alpha =  1.7 \pm 0.1$ and $\chi^2/dof = 65.36/65 \sim 1.005$. 

{\em Chandra} observed the afterglow of GRB~080207 on the 16th of February 2008. The afterglow 
was placed on the AXAF CCD Imaging Spectrometer (ACIS) S-3 (back illuminated) chip and Very Faint (VF) mode 
employed to
enable better rejection of background events. The standard cleaned 
event files were utilised, but filtered to
the energy range of  0.5-7~keV (largely to reduce background 
events and better isolate the afterglow). The afterglow was detected at a position of 
RA=13$^h$ 50$^m$ 02.98$^s$ , Dec = 07$^\circ$ 30$^{\prime}$ 07.4$^{\prime\prime}$ (J2000) with a 0.5~arcsec error circle. 
The background subtracted count rate of the afterglow in this band was
found to be $5.3  \times 10^{-4}$~s$^{-1}$. There are insufficient counts in the 
image to obtain a spectrum directly, however, by assuming 
the same spectral index as measured in the {\em Swift} PC mode data this 
implies a flux 
of $3.8 \times 10^{-15}$~erg~s$^{-1}$ cm$^{-2}$  in the 0.3-10~keV band 
equivalent to \emph{Swift}/XRT, 
and is consistent to $\sim 1 \sigma$ with the extrapolation of the earlier 
X-ray lightcurve -- 
indicating that any jet-break has jet to occur 8 days post burst. Alternatively
the jet break could have occurred earlier than the onset of the PC mode observations ($\sim 5000$~s),
although this is unusual.

\subsubsection{Optical}
Deep optical observations of GRB~080207 were pursued by several groups roughly 
12 hours
after the GRB and include observations by 2 to 8 metre class telescopes in both 
the optical and nIR. None of these
observations yielded any afterglow candidates to deep limits. 
\cite{2008GCN..7279....1K} report deep optical limits from
the Gamma-Ray burst Optical/Near-infrared Detector
(GRON)D: g$^{\prime}>$23.9, r$^{\prime}>$23.8, i$^{\prime}>$23.5 and z$^{\prime}>$22.8, nIR limits from the
Very Large Telescope (VLT) are 
reported by \cite{2008GCN..7293....1F} as J$>$23.5, H$>$22.8 and K$>$21.5.

These limits are amongst the deepest obtained for emission from any GRB at 
moderate times
after the burst ($\sim 12$ hours), and were obtained across the optical and nIR 
waveband by the dual use of multiple 8m telescopes. The deep limits in both
the optical and the IR rule out colours similar to that of high-$z$ GRBs like 
050814 \citep{2006A&A...447..897J}, 
050904 \citep{2006Natur.440..184K,2006Natur.440..181H}, 080913 
\citep{2009ApJ...693.1610G,2010A&A...510A.105P,2010A&A...512L...3P},
090423 \citep{2009Natur.461.1254T,2009Natur.461.1258S}, 090429B \citep{2011ApJ...736....7C}, and also very red 
colours due to extinction as have been observed in a handful of bursts
\citep[e.g.][]{2006ApJ...647..471L,2007ApJ...669.1098R,2008ApJ...681..453J,2008MNRAS.388.1743T}. 

\subsection{Astrometry}
To locate the X-ray afterglow precisely on the deep host galaxy images, relative astrometry was performed 
between the {\em Chandra} frames and those obtained at the 
VLT (see Section \ref{sec:host}). Sources located in the {\em Chandra} frame were centroided by fitting Gaussian 
profiles to their light distributions. These were then compared with the 
VLT Focal Reducer and low dispersion Spectrograph 
(FORS2)\footnote{Although tying directly to the {\em HST} images would have been preferable, this is unfeasible due to the small field of view
which contained too few sources in common.} 
frame (see section \ref{FORS2}), giving a total of 6 optical counterparts to X-ray sources in the 
optical image. An astrometric solution was computed with the IRAF task {\tt geomap}, 
which places the afterglow on the FORS2 frame with an accuracy of 0.45 arcsec. 
Subsequent relative astrometry between the FORS2 and {\em HST} WFPC2 and NICMOS 
frames was performed using 10 (WFPC2, F606W) and 7 (NICMOS, F160W) sources in common to each frame. 
The total error in the placement of the X-ray afterglow on the {\em HST} images is 
$\sim 0.5$ arcsec. We additionally placed the afterglow on our other optical/IR frames 
(other HST filters and instruments, Gemini and Spitzer observations) by performing relative
astrometry between NICMOS and those images, the resulting error on these transformations
is typically very small ($\sim 0.1$ arcseconds) and does not contribute significantly to the
overall positional error budget. 

\subsection{Host galaxy}
\label{sec:host}
At the location of the X-ray afterglow we clearly find a extremely red host galaxy, with $g=27.41 \pm 0.3$ and $Ks=21.74 \pm 0.13$ (AB-mag, see 
below for more details). The probability of a chance alignment of a $g \sim 27.5$ galaxy is moderate, even within our
0.5 arcsecond error circle, with $P_{chance} \sim 0.1$ following \cite{2002AJ....123.1111B,2007MNRAS.378.1439L}. However, 
the probability of an ERO is much lower, indeed, simply utilising the K-band number counts \citep{2008MNRAS.383.1366C} and not accounting
for the colours, would imply a probability of $P_{chance} \leq 1\%$. Hence we identify this galaxy as the host of GRB 080207\footnote{We note that while this paper was
in review, a separate paper by \cite{2011ApJ...736L..36H} has appeared. Our precise position confirms their
host identification, and their independent discovery of the ERO host.}. 
We have acquired deep observations of the host galaxy
host galaxy in 19 bands ranging from observed frame optical B-band to sub-mm 
$850\,\rm{\mu m}$. The host galaxy is faint or undetected in the optical and bright at longer wavelengths, 
indicating very red colours not usually associated with GRB hosts.
Various images of the field of the host galaxy are displayed in Figure \ref{fig:ds9}. The XRT position (large green circle)
is unable to uniquely determine the host, while the improved {\it Chandra} position (small red circle)
intersects three small knots with similar colours, which will be assumed to belong to the host galaxy system.

\subsubsection{Hubble Space Telescope}
The X-ray position of GRB~080207 was observed by the {\it Hubble Space Telescope} ({\it HST})
using both the Wide Field and Planetary Camera-2 
(WFPC2) in the F606W, F702W and F814W filters, 
the Near Infrared Camera and Multi-Object Spectrometer
(NICMOS) with the NIC3 
camera and F160W
filter (H-band) and 
the Wide Field Camera-3 (WFC3) with the F110W filter. Details of the individual 
observations are reported in Table~\ref{observations}.

The WFPC2 data were retrieved from the archive with ``On-The-Fly'' processing. 
The individual images were
then cosmic ray rejected, shifted and combined via {\tt multidrizzle} 
to produce a final image with a scale of 0.06~arcsec per pixel (roughly
2/3 of the native pixel size). 

NICMOS images were cleaned for quadrant dependent residual bias levels 
(pedestal effect) using {\tt pedsky} and subsequently processed 
through {\tt multidrizzle} onto an output grid with pixel size 0.1~arcsec. WFC3 
observations were obtained with a standard 4-point box dither pattern, and also 
combined
via {\tt multidrizzle}, with the native pixel size unchanged (0.13~arcsec). 

There is no evidence for host galaxy emission in any of the WFPC2 observations. 
However, the F160W observations clearly show evidence for a host galaxy at the location of 
the X-ray afterglow
of GRB~080207. Point source limits for objects at the location of GRB~080207 
in the
WFPC2 images are F606W = 26.8, F702W=27.2, F814W=27.0 (all 3$\sigma$ AB 
magnitude limits). 
However, the galaxy is clearly extended in the F160W observations, hence we
attempt to derive more realistic
limits using apertures equal to the half light radius of the galaxy as measured 
in the F160W observations (0.4 arcsec), and then assumed this accounted for
only 50\% of the total galaxy light, hence brightening the limits by a factor of two. 
In practice the true limiting magnitude depends on the distribution of
light within the galaxy, where bright knots of emission could often be seen, even if 
low surface brightness areas were missed. However, these limits are broadly in agreement with the 
magnitude limits obtained by 
populating the images with fake sources of half light radii equal to that of 
the host, and subsequently 
attempting to recover them via {\tt SExtractor} \citep{1996A&AS..117..393B}. The resulting limits are F606W=25.4, 
F702W=25.65, F814W=25.02. 
(See also Table~\ref{observations})

\subsubsection{Ground based host observations}
In addition to the optical and nIR observation with {\it HST}, deep 
imaging of the host galaxy was obtained with the VLT, Gemini and Keck.

\label{FORS2} The VLT R-band observation was carried out on the 1st of April 2008, using FORS2. 
In this image, the host galaxy remains 
undetected to a limit $\sim 25.65$. Although the R-band limit is affected
by blending with a neighbouring source, the limiting magnitude is broadly 
consistent with that from {\it HST}.

The Gemini imaging was obtained with the Gemini Multi-Object Spectrograph (GMOS)
in the z-band, and the Near Infra-Red Imager and spectrometer (NIRI) in J and K.
The seeing in the z-band 
observations was very good ($\sim 0.5$~arcsec), but was poorer for the J and K 
band ($\sim 0.9$~arcsec).
These observations were reduced in the standard fashion under IRAF. The host is 
detected in each of these observations, although
only with marginal significance in the J-band observations. Photometry of the 
host galaxy
was performed relative to Sloan Digital Sky Survey (SDSS) observations of the field for the z-band 
observations, 
and in comparison to Two-Micron All Sky Survey (2MASS) for the J and K. 

The Keck observations were performed with the Low Resolution Imaging Spectrometer
(LRIS) in the g- and I-bands.
The images were reduced with standard IRAF techniques and zero magnitudes were calibrated relative to SDSS
stars in the field. We note that both the g- and I-bands are deeper than the {\it HST} and Gemini optical observations, resulting
in a detection of the host at low wavelength indicating redshift $z < 2.8$.
The Ks observations provided
a factor $\sim 2$ better signal-to-noise than the Gemini observations in the K-band, and 
flux consistent within $1\sigma$. The Ks-band is calibrated using sources in the field common with the
Gemini frame.
See Table~\ref{observations} for a full summary of all observation details and results.

\subsubsection{Spitzer}
The host of GRB~080207 was also observed by the {\em Spitzer Space Telescope}, 
utilising both the Infra-Red Array Camera (IRAC) in
all 4 bands (3.6, 4.5, 5.8 and 8.0 \micron) and with 
the Multiband Imaging Photometer (MIPS) at 24~\micron. The 
host is clearly detected 
in all IRAC and MIPS bands, indicating significant nIR and mIR emission, 
possibly suggesting a massive and dusty host respectively.
The clear detections in these bands are in contrast to the majority of GRB 
hosts which are undetected (or very weakly 
detected) in similar observations 
\citep[e.g.][]{2006ApJ...642..636L,2008arXiv0803.2235C}. As the host is 
unresolved at the resolution of {\em Spitzer}, 
photometry of host was performed on the standard post-basic calibrated data (BCD) mosaics, utilising 
small apertures 
($2.4$ and $7.4$~arcsec for IRAC and MIPS respectively) and applying tabulated aperture corrections and 
zeropoints. The resulting magnitudes
are shown in Table~\ref{observations}.

\subsubsection{SCUBA2}

As a part of the early ``shared risk" operations 
with SCUBA2 \citep{2006SPIE.6275E..45H,2008ASPC..394..450E} on the JCMT,  we obtained $\sim$43 minutes of observations in
the $450\,\rm{\mu m}$ and $850\,\rm{\mu m}$  bands during the nights 2010-02-25,2010-02-26 
and 2010-03-12. 
The imaging was carried out in the SCAN mode with a DAISY scanning 
pattern. 
The data were reduced using the STARLINK module {\tt SMURF}, running makemap
 in the 
iterative mode\footnote{i.e. iteratively fitting detector signal and background noise.}
 to map the SCAN data into a sky image with a pixel scale of 3~arcsec \citep[e.g.][]{2010arXiv1011.5876J}. 
The sky maps are flux calibrated relative to the sub-mm flux of CRL618 which was 
observed during
the same nights as the science observations \citep[e.g.][]{2010SPIE.7741E..54D}. Before the maps for all nights were 
co-added,  astrometric corrections were applied
as determined by separate observations of pointing sources obtained during each night.
We performed aperture photometry in the $450\,\rm{\mu m}$ and $850\,\rm{\mu m}$ bands 
respectively - measuring fluxes $23037 \pm 17740\,\rm{\mu Jy}$ and  $2529  \pm 4374  
\,\rm{\mu Jy}$ respectively, although the host is undetected. Using blank apertures 
on the map we estimate  $3\sigma$ limiting
magnitudes of $12.1$ and $13.6$ (AB magnitudes) in the  $450\,\rm{\mu m}$ $850\,\rm{\mu m}$ 
bands respectively, which offer only weak constraints on the sub-mm emission.

\section{Afterglow properties}
The X-ray spectrum exhibits apparent absorption significantly in excess of the 
Galactic value. The preferred zero redshift model results in
$N_H \sim 75 \times 10^{20} \mathrm{cm^{-2}}$ (c.f. total Galactic $N_H$ column 
$\sim 1.94 \times 10^{20} \mathrm{cm^{-2}}$) 
with $\chi^2 / dof$ = 125/153. Attempting to fit a broken power law with fixed 
$\Delta \beta =0.5$, e.g. assuming the spectral turn-over to be influenced by a 
cooling break in the X-ray band, 
results in significantly worse fits with $\chi^2 / dof$ = 168/152, and 36/29 
respectively for PC and WT mode data, 
suggesting that excess $N_H$ is the most likely explanation for the observed 
spectrum. 
 
\cite{2007AJ....133.2216G} suggest that the X-ray measured $N_H$ column can be 
used to limit the redshift by
\begin{equation}
\log{(1+z)} < 1.3 - 0.5 \,\log_{10}{(1 + \Delta N_H)},
\end{equation}
where $\Delta N_H$ is the difference between Galactic and observed $N_H$ values 
in units of $10^{20}$~cm$^{-2}$, fitted at zero redshift. 
This would suggest that GRB~080207 originates from $z<1.3$. Interestingly the 
only GRB in the sample of \cite{2007AJ....133.2216G} to be found with a higher
$N_H$ than GRB~080207 is GRB~051022, whose optical afterglow was also markedly 
suppressed \citep{2007ApJ...669.1098R, 2007AA...475..101C}. Indeed, although it is commonly very 
difficult to assess 
the redshifts for dark GRBs it is occasionally possible to pinpoint redshifts 
for bursts
whose optical afterglows are somewhat suppressed, and are invisible to UVOT, 
but are still visible to deep ground based optical 
observations.  In these cases the measured (rest frame) column densities are 
apparently higher than
those for the GRBs with very bright optical afterglows \citep{2007MNRAS.377..273S}. 

Assuming that GRB~080207 is {\it not} limited to $z<1.3$, we fit the X-ray 
spectrum with single power law model absorbed 
by the Galactic $N_H$ column {\it and} an absorber redshifted to $z=1.74$ as 
suggested by out photometric redshift 
solutions for the host (see section 4.1). This model suggests an X-ray spectral 
slope $\beta=1.34^{+0.17}_{-0.16}$ and a significantly higher $N_H$ column than 
the zero redshift case
with $N_H = 679^{+125}_{-114} \times 10^{20} \mathrm{cm^{-2}}$. This makes this 
one of the highest measured restframe $N_H$ column of any GRB host yet.

Extrapolating the X-ray power law to optical/nIR frequencies and re-normalising 
the integrated flux to be consistent to the 11 hour post burst flux 
suggested by the lightcurve, reveals the optical/nIR flux limits are fainter 
than expected. The
X-ray-to-optical spectral slope is estimated to be $\beta_{OX}<0.3$ and thus this burst fulfil the 
criteria for dark bursts of \cite{2004ApJ...617L..21J} (and also
fulfils the dark criterion by \cite{2009ApJ...699.1087V} since $0.3 < 
\beta_X-0.5$).  
To evaluate an optical extinction that explains the optical darkness of this 
burst, we adopt extinction curves derived for
the Milky Way (MW)  \citep{1979MNRAS.187P..73S} with $R_V=3.1$; the 
Small Magellanic Cloud (SMC) \citep[][]{1984A&A...132..389P} with $R_V=2.72$; typical starburst galaxies (SB) 
\cite{2000ApJ...533..682C} with $R_V$=4.05; and 
the afterglow of the dark GRB~080607 \citep{2010arXiv1009.0004P}.

\begin{figure}
  \begin{center}
    \includegraphics[scale=0.4]{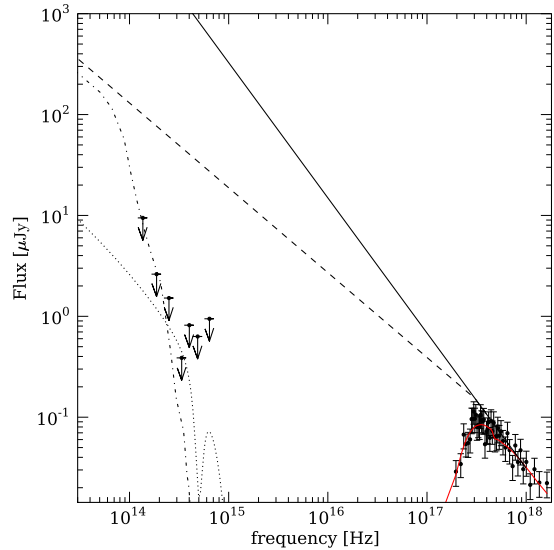}
    \caption[GRB~080207 afterglow SED]{The afterglow spectral energy distribution $\sim 11$ hours post burst, ranging  
    from nIR to X-ray frequencies. 
    The solid red line shows the X-ray model fitted with redshift $z=1.74$, 
    the solid black line is the X-ray power law extrapolated without a 
    spectral break and the dashed line with a $\Delta \beta = 0.5$ cooling
    break. 
    The power law and spectral break model is shown absorbed in the restframe 
    by a Milky Way reddening law with $A_V=2.6$ (dotted line),
    and by a SMC law with $A_V=3.7$ (dash-dotted line).}
    \label{fig:afterglow_sed}
  \end{center}
\end{figure}

The afterglow model SED is reddened after extrapolating the X-ray into the optical-nIR 
regime, and after introducing a cooling break with $\Delta \beta=0.5$ 
short-wards of the XRT band (Figure~\ref{fig:afterglow_sed}). By requiring that the
absorbed extrapolation falls below the detection limits, at the redshift 
$z=1.74$ a restframe line of sight 
extinctions in excess of $2.6$ (MW), $3.7$ (SMC), $4.1$ (SB) and $3.4$ (GRB~080607) 
magnitudes is found. These all suggest that the optical extinction is indeed also very high 
compared to the bulk GRB population, but that the dust-to-gas ratio is comparable 
to that found in other hosts \citep[e.g.][]{2010MNRAS.401.2773S,2009AJ....138.1690P}.
A summary of derived afterglow properties can be found in Table~\ref{tab:afterglow}.
\begin{table}
  \begin{center}
    \begin{tabular}{l|l}           
   \multicolumn{2}{c}{\bf Afterglow properties }\\ 
      \hline
      Ra, Dec(J2000)               & 13:50:02.98, +07:30:07.4                          \\
      Errorbox                     & 0.5 arcsec                                       \\
      \hline
      $\chi^2/dof$ (spectral fit)  & $48.49/48 \sim 1.01$                              \\
      $\beta$                      & $1.34^{+0.17}_{-0.16} $                             \\
      $N_H$                        & $679^{+125}_{-114} \times 10^{20} \mathrm{cm^{-2}} $ \\
      $A_V$ (MW law)               & $\geq 2.6$                                          \\
      $A_V$ (GRB~080607 law)      & $\geq 3.4$                                          \\
      $A_V$ (SMC law)              & $\geq 3.7$                                          \\
      $A_V$ (SB law)               & $\geq 4.1$                                            \\
      \hline
      $\chi^2/dof$ (lightcurve)    & $65.78/66 \sim 1.00$                                \\
      $\alpha$                     & $1.7 \pm 0.1$                                       \\
      \hline
      \hline
    \end{tabular}
  \end{center}
  \caption[GRB~080207 spectral fit parameters]{\emph{Chandra} X-ray position and fitted parameters for the 
    afterglows analysis. The quoted hydrogen column and extinction are calculated in the restframe of the 
    hosts photometric redshift ($z_{phot}=1.74$). 
  }
  \label{tab:afterglow}
\end{table}

\section{Host galaxy properties}
The g-band detection of the host galaxy suggests that it lies below $z \sim 4$.
Coupled with the relatively bright magnitudes in the nIR to mIR, and the red colours 
across the whole of the wavelength range, rather than a sharp break in the optical and a flat 
SED in the optical - nIR, the favoured interpretation is that of a dusty sightline. 
This is also strongly indicated by the detection of the host galaxy at 24~\micron, and although the SCUBA2 
limits are not deep enough to offer any significant constraints, they are fully consistent with 
sub-mm dust emission at the photometric redshift $z \sim 1.7$ we derive in Section \ref{sec:photoz}.

The observed lower limit on the colour of $R-K>5.4$ (equivalent to $R-K>3.7$ in AB magnitudes) 
 is one of the reddest GRB hosts yet discovered, and indicates that, at least in the case of GRB~080207, 
the environment is markedly different to that of optically bright bursts.
The high resolution imaging acquired by the WFC3 on {\em HST} resolves the 
large scale structure of the host, which is displaying
an irregular morphology,  
suggesting a merging or disturbed system, possibly crossed by dust lanes.

\begin{figure*}
  \begin{center}
    \includegraphics[scale=0.45]{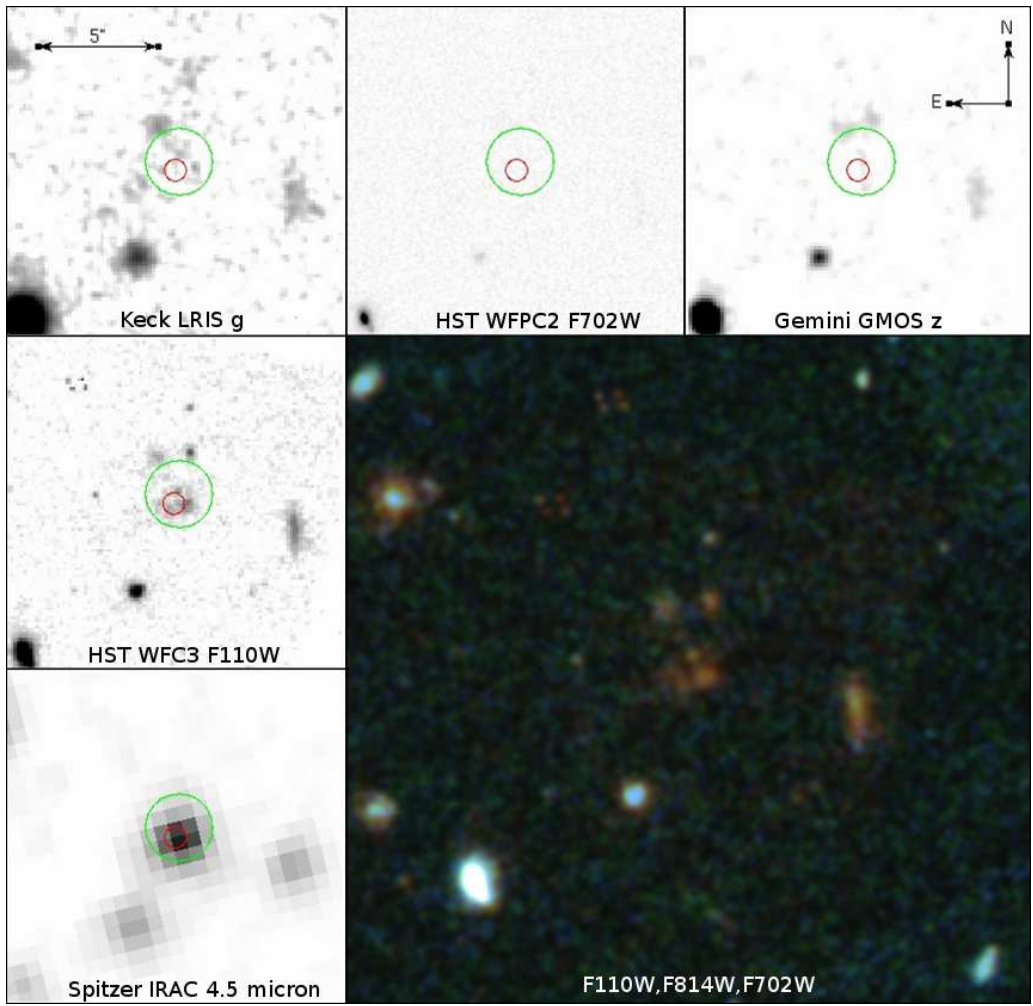}
    \caption[Five band mosaic image of the field of GRB~080207]{Five band mosaic 
      image of the field of GRB~080207 including its host galaxy (top and left panels). The red circle marks 
      the {\it Chandra} X-ray position and errorbox, the green circle show the {\it Swift}/XRT position and error-box. The host
      is faint or undetected in the optical but shows strong emission in nIR and longer wavelengths. The large lower right panel
    shows a 3 filter false colour image showing the extremely red host galaxy in the centre and a number of other red galaxies also in the field.}
    \label{fig:ds9}
  \end{center}
\end{figure*}

In the following section we will discuss the photometric redshift solutions and the 
restframe properties which it implies.
The 19 bands covered by photometry are presented in Table \ref{observations} 
and a four band mosaic image in Figure~\ref{fig:ds9}
shows the host going from non-detected in the visual, to faint in z-band to 
strong detections in nIR J-band and IR $4.5\,\rm{\mu m}$.
In the following we have assumed a $\Lambda CDM$ cosmology with 
$\Omega_M=0.27$, $\Omega_{\Lambda}=0.73$ and $H_0=71\,{\rm km\,s^{-1} Mpc^{-1}}$.

\begin{table*}
  \begin{center}
    \begin{tabular}{llllll}
      \multicolumn{6}{c}{\bf Host observation log} \\ 
      \hline
      Date                & Instrument         & Filter      & Exp.Time (s)& Magnitude (AB)    & flux ($\rm{\mu Jy}$) \\
      \hline
      2009-02-19          & Keck/LRIS          & g           & 1640      &  27.41 $\pm$ 0.3  & 0.04   $\pm$ 0.01        \\  
      2008-03-18          & {\em HST}/WFPC2    & F606W       & 1600      &  $>25.4$          & 0.16   $\pm$ 0.10       \\
      2008-04-01          & VLT/FORS2          & R           & 2000      & $>25.65^{1}$      & 0.14   $\pm$ 0.07        \\
      2009-03-21          & {\em HST}/WFPC2    & F702W       & 3600      & $>25.65$          & 0.2    $\pm$ 0.08        \\
      2009-02-19          & Keck/LRIS          & I           & 1500      & 25.84$\pm$ 0.29   & 0.17   $\pm$ 0.05        \\  
      2009-03-20          & {\em HST}/WFPC2    & F814W       & 3300      & $>25.03$          & 0.38   $\pm$ 0.13        \\
      2009-02-24          & Gemini/GMOS        & z           & 1260      & 25.02 $\pm$ 0.25  & 0.18   $\pm$ 0.05         \\
      2009-02-19          & Gemini/NIRI        & J           & 2880      & 23.87 $\pm$ 0.31  & 1.06   $\pm$ 0.35        \\
      2009-12-10          & {\em HST}/WFC3     & F110W       & 2400      & 23.32 $\pm$ 0.09  & 1.75   $\pm$ 0.17         \\
      2008-04-05          & {\em HST}/NICMOS   & F160W       & 2560      & 23.04 $\pm$ 0.14  & 2.27   $\pm$ 0.34         \\ 
      2009-02-19          & Gemini/NIRI        & K-prime     & 2880      & 21.94 $\pm$ 0.24  & 6.25   $\pm$ 1.62        \\ 
      2009-05-31          & Keck/NIRC          & K-short     & 1500      & 21.74 $\pm$ 0.13  & 7.52   $\pm$ 0.93          \\   
      2009-03-20          & {\em Spitzer}/IRAC & 3.6 $\rm{\mu m}$ & 1600 & 20.81 $\pm$ 0.04  & 17.7   $\pm$ 0.76         \\
      2009-03-20          & {\em Spitzer}/IRAC & 4.5 $\rm{\mu m}$ & 1600 & 20.67 $\pm$ 0.03  & 20.14  $\pm$ 0.65          \\
      2009-03-20          & {\em Spitzer}/IRAC & 5.8 $\rm{\mu m}$ & 1600 & 20.21 $\pm$ 0.13  & 30.76  $\pm$ 4.32         \\
      2009-03-20          & {\em Spitzer}/IRAC & 8.0 $\rm{\mu m}$ & 1600 & 20.63 $\pm$ 0.19  & 20.89  $\pm$ 4.29         \\
      2008-07-31          & {\em Spitzer}/MIPS & 24  $\rm{\mu m}$ & 5407 & 18.50 $\pm$ 0.20  & 148.59 $\pm$ 32.1         \\
      2010-02-25,26,03-12 & JCMT/SCUBA2        & 450 $\rm{\mu m}$ & 2616 & $>12.1 $          & $ 23040 \pm 17740 $        \\
      2010-02-25,26,03-12 & JCMT/SCUBA2        & 850 $\rm{\mu m}$ & 2616 & $>13.6 $          & $ 2530 \pm 4370 $        \\     
      \hline
      \hline
    \end{tabular}
  \end{center}
  \caption[Photometric observations of the GRB~080207 host galaxy]{Photometric 
    observations of the GRB~080207 host galaxy as part of this work. Magnitude are
    in the AB system. $^{1}$indicates blending with a nearby source affects 
    the limiting magnitude. Limits in the magnitude column are $3\sigma$ estimated 
    from half-light radius apertures (WFPC2) or point source limits (SCUBA2). In 
    the flux column, the actual flux measured in the aperture also in the cases of 
    non-detections, are reported.}
  \label{observations}
\end{table*}

\subsection{Photometric redshift}
\label{sec:photoz}
The 19-band observations cover a broad
wavelength range from optical to sub-mm, and should allow a well constrained 
photometric redshift to be determined, and estimates of the physical properties (e.g. mass and star formation rate) of the host galaxy 
to be made without relying on extrapolating an assumed spectral shape.
To enable detailed and accurate modelling of a system that could possibly contain both a young and starbursting
stellar population and an older, redder component we find that allowing for a linear combination
of two templates provide a significantly better fit than only using a single template. 
Hence, to simultaneously fix the photometric redshift and the full restframe spectral energy distribution,
we fitted a linear combination of two templates: one coming from a set of detailed optical templates including
models described in \cite{1980ApJS...43..393C} and \cite{1993ApJ...405..538B}; and the second set of templates \citep[described by][]{2007A&A...461..445S}
containing galaxies with significant amounts of dust increasing their IR and sub-mm luminosities by reprocessing the
UV and optical light. 
Furthermore, we fitted the reddening of the first set of models by assuming a \cite{2000ApJ...533..682C}
reddening law. The dusty templates in the second set already include a dust screen model, and are not reddened any further.
In total this comprises 6 free parameters (redshift, $A_V$, two templates and two normalisation constants.), and for 19 photometry data points gives $dof=19-6=13$ degrees of freedom.

\begin{figure}
  \begin{center}
    \includegraphics[scale=0.4]{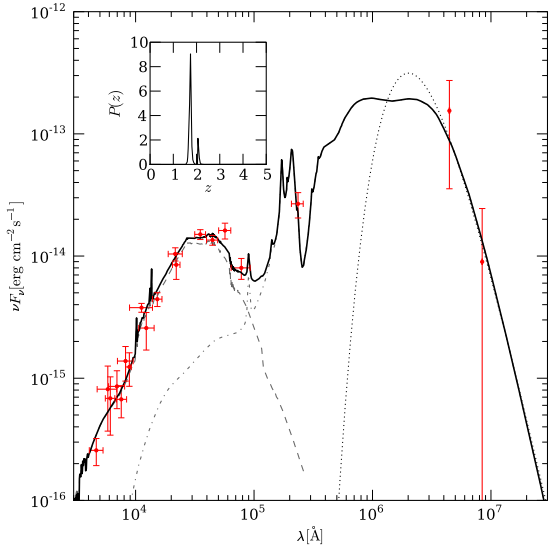}
    \caption[GRB~080207 host galaxy SED fit]{The host galaxy spectral 
energy distribution and photometric redshift solutions at $z_{phot} = 1.74^{0.06}_{-0.05}$. 
The wavelength scale is in the observer frame. The thick solid line 
shows the composite template model with the dashed, and dash-dotted lines 
showing the individual components. The dotted line is purely thermal 
emission from $\sim 7 \times 10^8$ M$_{\odot}$ dust at $\sim$ 45~K. 
The inset figure shows the probability distribution 
as a function of redshift. Errors are $1 \sigma$.}
    \label{fig:host_sed}
  \end{center}
\end{figure}

Fitting the available photometry, including measured fluxes for the non detections, and allowing
both redshift and host absorption to vary as free parameters 
\citep[see][]{2010MNRAS.405...57S} yields a primary
photometric redshift solution of $z=1.74^{+0.06}_{-0.05}$ with $\chi^2/dof =19.37/13 \sim 1.49$, 
shown in Figure~\ref{fig:host_sed}. The redshift error is the central $1\sigma$ interval, i.e 
the integrated probabilities above and below the interval are both $(1-0.683)/2$.  This confidence
interval is relatively narrow, but somewhat broader at 2$\sigma$, providing $z=1.74^{+0.34}_{-0.18}$.
This result is broadly consistent with an independently derived solution with {\tt Hyperz} 
\citep[][]{2000A&A...363..476B} using
only the optical and nIR photometry. It is also broadly consistent at the $\sim1.5\sigma$ level,
with the $z = 2.2^{+0.2}_{-0.3}$ obtained
by Hunt et al. 2011, using a smaller dataset. We do not attempt to increase the errors due to possible
systematic offsets between different instruments, however note that this would not change our photometric
redshift, but (for modest additional errors) would simply slightly increase the confidence ranges.
It is also worth noting that a higher redshift than provided by the best fit would further 
increase the restframe hydrogen column derived from
the X-ray spectrum, e.g. $\sim 10\%$ higher at $z=2.2$. A significantly higher 
solution is effectively ruled out by the g-band detection.

\subsection{Restframe properties of the host}
The restframe properties of the host galaxy as derived from these fits are 
shown in Table~\ref{tab:host}.
We estimated physical galactic properties from the restframe k-corrected and 
extinction corrected spectral energy distribution. Stellar mass content is 
estimated from the restframe
K band absolute magnitude \citep{2009ApJ...691..182S}, corresponding to between IRAC $5.8$ and 8~\micron\ at $z=1.74$. 
For the star formation rate we make two estimates, one based on the U-band luminosity 
\citep{1998ApJ...507..155C}
and one based on the far-InfraRed (fIR) luminosity \citep{1998ARA&A..36..189K}.  
The host is massive and rapidly star forming - assuming that the fIR traces the 
true star-formation rate (SFR) more accurately than the U band. Placing  it on the $\frac{SFR}{M_*}$ 
vs $M_*$ plane compared to the bulk GRB hosting galaxy population 
\citep[e.g.][]{2006ApJ...653L..85C,2008arXiv0803.2235C,2009ApJ...691..182S}
suggests that it is one of the most massive and most actively star forming GRB 
hosts to date. 
From the SED model we estimate a restframe fIR luminosity $L_{fIR} \sim 3 \times 10^{12} L_{\odot}$
suggesting that GRB~080207 is one of few bursts with a ULIRG host \citep[][]{2008ApJ...672..817M}. 
However it should be noted that the ULIRG classification rests mainly on the 24~\micron\ MIPS
detection, and while the SCUBA2 limits are consistent, they are also too bright to offer
significant constraints on the fIR nature of the SED.

Comparing the host with the luminosity function at $z \sim 2$ 
\citep[e.g.][]{2005ApJ...631..126D,2007ApJ...654..172D} suggests that it  
is comparable to the characteristic luminosity in the B-band; $L_B \sim 1.3 
L^*_B$, in contrast to the typically under-luminous properties of optically 
bright GRB selected samples.

In particular, it is clear that the host extinction in this case is high in 
comparison to the bulk GRB population - the dominant model in the optical
has an  $A_V \sim 1.9$ while the second component has a total of $\sim 100$ 
magnitudes of extinction from core to surface (see \cite{2007A&A...461..445S} 
for a description of their dust model) -
suggestive of a major dust content within the host. 
Although a $3\sigma$ detection is lacking
from SCUBA2, we estimate a $3\sigma$ upper limit of the dust mass as $\sim 1.2 - 1.4 \times 10^9$ M$_{\odot}$ 
assuming a dust temperature of 45~K \citep[e.g.][]{2008ApJ...672..817M}, and also note that 
a lower temperature would increase the necessary dust mass.
The possibility of significant dust content is in contrast to the majority of GRB host 
galaxies, whose photometry
suggests relatively little dust 
\citep[e.g.][]{2009ApJ...691..182S,2004MNRAS.352.1073T}, indeed it is more 
similar to that commonly 
found in sub-mm selected galaxies \citep[e.g.][]{2010A&A...514A..67M}.
However, it should be noted that these studies have mainly concerned host samples
optically selected, and \emph{may not} be representative of the true population.
\begin{table}
  \begin{center}
    \begin{tabular}{ll}
      \multicolumn{2}{c}{\bf Host restframe properties} \\ 
      \hline
      $z_{phot}$             & $1.74^{+0.06}_{-0.05}$\\
      $\chi^2/dof$         &19.37/13 $\sim$ 1.49  \\
      \hline
      $A_V$                 &$\sim 1.9$       \\
      $M_U$                 &$-20.29 \pm 0.04$     \\           
      $M_B$                 &$-20.99 \pm 0.04$     \\           
      $M_V$                 &$-21.86 \pm 0.04$     \\           
      $M_K$                 &$-23.89 \pm 0.04$     \\           
      $L_{fIR}$             &$2.4 \pm 0.09 \times 10^{12} L_{\odot}$ \\
      $\log_{10}{(M_{\star}/\Mo)}$     &$11.05 \pm 0.02 $    \\
      $SFR_U $              &$40.7 \pm 1.6  \Mo/\mathrm{yr}$    \\
      $SFR_{fIR} $          & $416 \pm 17.0$   $\Mo/\mathrm{yr}$   \\
      \hline
      \hline
    \end{tabular}
  \end{center}
  \caption[GRB~080207 host properties]{Restframe properties of the hosts SED 
    template fit. Absolute magnitudes are not corrected for host extinction. Stellar mass and star formation
    rates are corrected for a host internal extinction of $A_V=1.9$. The quoted errors are $1 \sigma$ statistical errors on the
    best fit template.}
  \label{tab:host}
\end{table}

\section{Discussion}
\subsection{Implications for dark GRBs}
GRB 080207 (see also Hunt et al. 2011) is one of very few GRB hosts which can be classified as an ERO. The 
other examples GRBs 030115 \citep{2006ApJ...647..471L} and 020127 
\citep{2007ApJ...660..504B} also
host bursts which were dark, or showed significant extinction in their 
afterglow lightcurve. 
Several other bursts also show very red colours in their afterglows, indicating 
significant extinction along the line of sight 
\citep[e.g.][]{2008MNRAS.388.1743T}, however
at least in some cases where the afterglow is unusually red, observations of 
the host galaxies do not reveal exclusively
red colours 
\citep[e.g.][]{2003A&A...400..127G,2003A&A...409..123G,2007ApJ...669.1098R,2008ApJ...681..453J,2009AJ....138.1690P,2001ApJ...562..654D},
although there is an apparent trend for the dark GRB host population to include 
much redder galaxies 
than that of the optically bright population 
\citep[e.g.][]{2010arXiv1003.3717H,2010arXiv1005.1257K}. 
Indeed, GRB hosts in general are very blue and typically sub-luminous 
\citep{2003A&A...400..499L,2004A&A...425..913C}, suggesting that only
a relatively small fraction of GRB selected star formation is obscured - at 
least so far as the bulk GRB hosting population
is represented by bursts with optically bright afterglows. 
Further the blue colours of the GRB hosts, and the relatively low detection
rate at long wavelength \citep[e.g.][]{2003ApJ...588...99B,2004MNRAS.352.1073T} 
in the pre-{\em Swift} sample suggest that few GRB
hosts are dusty systems, in contrast to sub-mm observations operating in
a similar redshift range, which suggest that the bulk of star formation is 
obscured, with a good fraction occurring in ULIRG-like galaxies 
\citep[][]{2005ApJ...622..772C,2010A&A...514A..67M}.

At first sight then it would appear that the complete set of galaxies hosting GRBs are very 
different from those of sub-mm galaxies, although the direct comparison is far from trivial \cite[e.g.][]{2004A&A...425L..33W}. 
Indeed, when comparing the rate of sub-mm detections with that expected under simple models 
of paucity, sub-mm bright GRB hosts are only marginally ($\sim 2\sigma$) below 
the expected values \citep[][]{2004MNRAS.352.1073T,2006ApJ...642..636L}. Though it 
should be noted that the sample of sub-mm observations of hosts is relatively 
small, and that this host sample 
had a median redshift $\sim 1.2$ compared to the median redshift of sub-mm 
galaxies $\sim 2.2$ \citep{2005ApJ...622..772C}

An alternative approach is to study the optical/IR properties of both GRB hosts 
and sub-mm galaxies.
The median I-K colour of sub-mm selected galaxies is I-K = 4.1 $\pm$ 0.2 
\citep{2004ApJ...616...71S}, much
redder than the general field population which has median I-K = 2.8 $\pm$ 0.1 
\citep{2004ApJ...616...71S}. In contrast the 
GRB population is typically very blue (if somewhat heterogeneously selected to 
date), 
with mean colours for optically bright bursts of I-K = 1.6 $\pm$ 0.3, based on 
the sample of 
\cite{2009ApJ...691..182S}, although a significant fraction of GRB hosts are 
undetected in 
deep K-band observations, implying at times even bluer colours. 

\begin{figure}
  \begin{center}
    \includegraphics[scale=0.4]{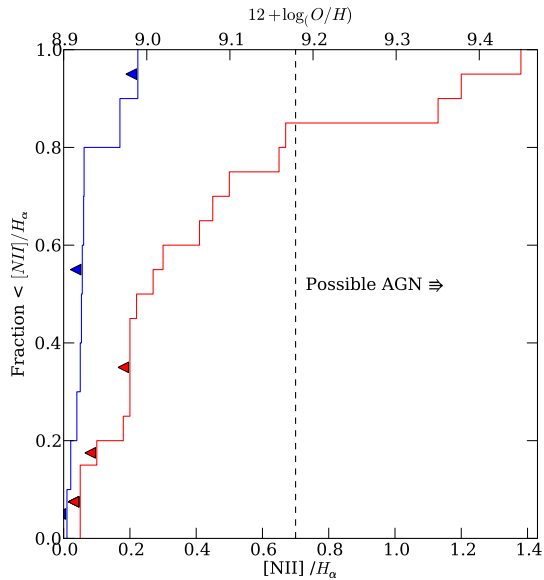}\\
    \caption[Metallicity distributions]{Cumulative distributions of the 
$[NII]/H\alpha$ ratio for low redshift ($z<0.7$), optically bright GRB hosts 
(blue) in comparison 
      to $z \sim 2$ sub-mm galaxies (red). Triangles indicate upper limit 
measurements. Sub-mm galaxies with $[NII]/H\alpha >0.7$ may have
    AGN contribution. All galaxies with H$\alpha$ restframe 
$FWHM < 1000$~km~s$^{-1}$ from \protect\cite{2004ApJ...617...64S} have been included.}
    \label{fig:met}
  \end{center}
\end{figure}

The mean ratio of [N {\sc ii}] / H$\alpha$ in sub-mm galaxies at $z \sim 2$ is 
of order 0.5 
based on deep IR spectroscopy \citep{2004ApJ...617...64S}, in contrast the 
(relatively local)
GRB hosts with the same measure yield  [N {\sc ii}] / H$\alpha$ $\sim 0.1$ 
\citep{2009ApJ...691..182S,2010arXiv1006.3560L}. This suggests 
that even at $z \sim 2$, where the universal metallicity may have dropped 
significantly,
sub-mm bright galaxies may not be the most promising locations for GRBs. 
Indeed, the
highest [N {\sc ii}]/H$\alpha$ ratio in the optically bright GRB sample of 
$\sim 0.2$ would only include
approximately $\sim 20$\% of the sub-mm sample of \cite{2004ApJ...617...64S} as 
shown in
Figure~\ref{fig:met}. Although few hosts of dark bursts have direct 
measurements of their metallicities,
making a direct comparison difficult, we note that the dark GRB~020819 has the 
highest measured [N {\sc ii}] / H$\alpha$ so far reported \citep{2010ApJ...712L..26L},
suggesting the corresponding distribution for dark bursts  includes metallicities 
at least $\sim \times 2$ higher.
Future observations of the  [N {\sc ii}]/H$\alpha$ ratio in GRB hosts at 
higher-$z$ 
(for example in the IR with X-shooter), should enable firm statistical 
statements to 
be made. In the meantime, we can discuss the host mass distribution 
which provides a rough proxy for the metallicity distribution.
%

\subsection{The mass distribution of dark burst hosts}
\begin{figure}
  \begin{center}
    \includegraphics[scale=0.4]{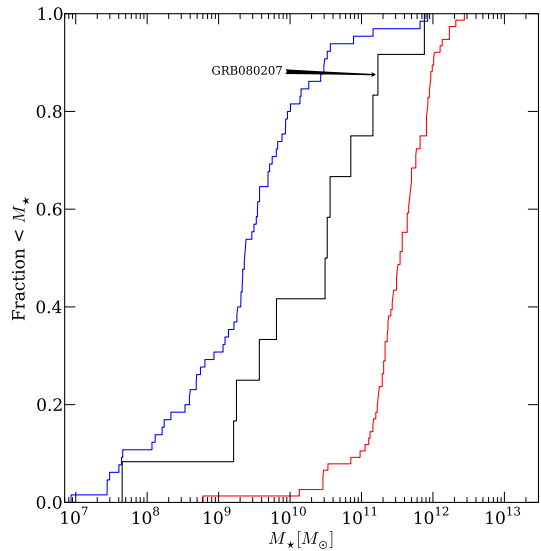}
    \caption[Dark bursts, host mass distribution]{Cumulative distribution of 
stellar mass in optically bright GRB host galaxies (blue line) and hosts of 
dark bursts (black line). 
      For a comparison we also show the distribution of stellar masses of the 
sub-mm galaxies (red line) calculated by 
\protect\citep[][]{2010A&A...514A..67M,2010ApJ...712..942M}
    }
    \label{fig:mstar}
  \end{center}
\end{figure}

In order to further understand the relations between the dark burst hosting 
galaxy population and  ULIRG / sub-mm like galaxies,
we compare the stellar mass distributions of sub-mm galaxies calculated by 
\cite{2010A&A...514A..67M,2010ApJ...712..942M} with the stellar masses
of dark burst hosts (see Table \ref{tab:darkhosts}) and the optically bright bursts to redshift $z < 4$.
We also estimate the sub-mm galaxy masses with our own SED fitting code, and 
note that results are consistent with the adopted values.
The cumulative mass distributions are shown in Figure~\ref{fig:mstar}.
While it is important to note that the host sample of dark GRBs consists of 
only 11 galaxies, the results clearly show that dark bursts are systematically 
hosted by
the most massive systems compared the optically bright GRBs. The formal 
probability that the samples of optically dark and optically bright bursts are 
drawn from the same population is given by the 
Kolmogorov-Smirnov (KS) test, where $P_{KS}=0.009$. 
The contrasting host masses between optically bright and dark bursts is 
particularly interesting as it lends further credibility to claims that samples
based primarily on bursts with optically detected afterglows could be severely 
inhibited by selection effects \citep[e.g.][]{2009ApJS..185..526F}. 
\begin{table}
  \begin{center}
    \begin{tabular}{l|lll}
      \hline
      GRB & $z$&$\log_{10}(M_*/\Mo)$ & Ref. (mass or photometry)\\
      \hline
      970828  & 0.958 & 9.57   & \cite{2010MNRAS.405...57S} \\
      000210  & 0.846 & 9.21   & \cite{2010MNRAS.405...57S} \\
      020819  & 0.41  & 10.52  & \cite{2010MNRAS.405...57S} \\
      050223  & 0.59  & 9.81   & \cite{2010MNRAS.405...57S} \\
      051022  & 0.807 & 10.49  & \cite{2010MNRAS.405...57S}\\
      060210  & 3.9   & 10.56  & \cite{2009AJ....138.1690P} \\  
      061126  & 1.16  & 11.16  & \cite{2010MNRAS.405...57S}\\
      061222  & 2.08  & 7.65   & \cite{2009AJ....138.1690P} \\
      080207  & 1.74  & 11.05  & this paper\\
      080325  & 2     & 10.85  & \cite{2010arXiv1003.3717H}\\
      080607  & 3.036 & 11.88  & \cite{2010arXiv1010.1002C}\\
      090417B & 0.3   & 9.25   & \cite{2010ApJ...717..223H}\\
      \hline
      \hline
    \end{tabular}
  \end{center}
  \caption[Stellar masses of dark burst hosts]{Stellar masses of all host 
galaxies of dark bursts available to date. Note that in the case of 
GRB~090417B we have supplemented the existing data with 
    additional photometry and derived new stellar mass estimates.}
  \label{tab:darkhosts}
\end{table}

Although we have not been able to reach a detection of the host sub-mm flux by 
SCUBA2, 
the number of GRB hosts with significant dust content can roughly estimated. Assuming that some 
fraction of dark bursts occur in obscured systems,
and also have similar dust to mass ratios -- we compare their stellar mass 
distributions in Figure~\ref{fig:mstar}. Roughly estimated, $\sim 90 \%$ of 
the dark burst hosts are more massive than the least massive sub-mm galaxy -- 
and hence under this simple argument one could expect 
a similar detection rate  of dark GRB hosts in the sub-mm at SCUBA sensitivity. 
Depending on the intrinsic mass function of the sub-mm population, 
even greater  detection rates could be plausible with SCUBA2 and even
with short integrations with ALMA when 
considering that the sub-mm galaxy sample in this comparison is flux-limited 
\citep{2005ApJ...622..772C}. In terms of physical properties 
of the dark burst hosts, this suggests that dark bursts are hosted 
predominantly by a very dust rich galaxy population. 

Given that GRBs trace (at best) a fraction of star formation, potentially even 
at moderately large
redshift it is surprising that attempts to transfer directly between GRB rate 
and star formation rate
produce even moderately consistent results 
\citep[e.g][]{2004AIPC..727..503P,2008ApJ...683L...5Y,2009ApJ...705L.104K}. 
Although the sample of dark bursts to date with detected and studied host 
galaxies is still small, the
emerging picture suggests that they indeed trace a different galaxy population 
than the optically bright sample.
 Certainly the host of GRB080207 is more akin to sub-mm or ULIRG galaxies than 
to the typical GRB hosts, 
suggesting that it is part of a subset of the GRB hosting galaxy population 
that trace star formation in more
massive, dusty and metal rich environments. 
In the face of the growing evidence that dark bursts can be hosted at higher 
metallicity than the bulk GRB population
studied today, it should be considered likely that GRBs can offer significant advantage over 
other methods to study the evolution
of the cosmic star formation history -- but only by paying due attention to 
sample selection effects and understanding the dark burst host population to avoid bias effects.

Although there is no direct measurement of the metallicity of the host of 
GRB~080207, the high stellar mass
is suggestive of a metal enriched environment -- again raising the question of 
what is the nature and metallicity dependence 
of GRB progenitors?
Considering the low metallicities typically associated with the bulk of the GRB 
hosts, we note that several
authors \cite[e.g.][]{2006MNRAS.372.1351L,2007AIPC..906...69D} have discussed 
tight binary systems as possible progenitors to GRBs 
in high metallicity environments.
While this would still require ongoing star formation and high mass stars, 
\cite{2010arXiv1005.0511H} report evidence for top-heavy IMFs in merging 
systems, increasing the likelihood of a GRB progenitor.
 
If the galaxy hosting GRB~080207 is undergoing a merger that further increased 
its rate of forming massive stars, and if a binary
progenitor is indeed possible at high metallicity - maybe massive and dust-rich 
galaxies are hosting a non-negligible fraction
of bursts. Although to which extent these conclusions can be generalised to 
other dark bursts is far from certain.

\section{Summary}
We have studied the afterglow of the dark GRB080207 from X-ray to nIR 
wavelengths and presented evidence of significant 
extinction in excess of at least 2.6 magnitudes (MW law) in the restframe visual as the 
cause of its optical-nIR darkness. The
high optical extinction is also echoed by the restframe hydrogen column which 
is the highest measured in any GRB
environment to date.
Lacking optical detection of the afterglow we have used observations 
of the X-ray afterglow at late time with {\it Chandra}, enabling an X-ray 
position to accurately identify the host galaxy. The ERO host spectral energy 
distribution has been studied in 19 bands from optical to sub-mm allowing us to 
estimate a photometric redshift $\sim 1.74$ and an average optical extinction of $A_V 
\sim 1.9$ in a massive galaxy. Furthermore, the host appears to be a ULIRG from its
fIR SED, with a high star formation rate as traced by the fIR light.
With a significant fraction of all bursts being classified as dark, and an 
increasing desire to utilise GRBs as high redshift
probes of the star formation evolution, the understanding of the nature of dark 
bursts should be highlighted.
This, and a number of other dark bursts in similar hosts should further 
encourage the study of dark bursts, their host environments
and how they relate to the evolving rate of star formation.

\section*{Acknowledgements}
We thank the referee for a careful review of this manuscript, which
has substantially improved the paper. 
We thank Harvey Tannenbaum and the CXO staff for their
help is securing rapid Chandra observations. KMS is grateful
to the University of Warwick for doctoral studentship. AJL
thank STFC for postdoctoral fellowship. NRT is grateful
to STFC for senior fellowship award. S.B.C. acknowledges
generous support from Gary and Cynthia Bengier and the
Richard and Rhoda Goldman fund. PJ acknowledges support
by a Marie Curie European Reintegration Grant within
the 7th European Community Framework Program, and a
Grant of Excellence from the Icelandic Research Fund. The
Dark Cosmology Centre is funded by the DNRF. Based on
observations made with the NASA/ESA Hubble Space Telescope,
obtained from the data archive at the Space Telescope
Institute. STScI is operated by the association of Universities
for Research in Astronomy, Inc. under the NASA contract
NAS 5-26555. The observations are part of proposal
number 11343. This work is based in part on observations made 
with the Spitzer Space Telescope, which is operated by the Jet 
Propulsion Laboratory, California Institute of Technology under a contract with NASA.
Based on observations made with ESO Telescopes
at the La Silla or Paranal Observatories under programme
177.A-0591. This research has made use of data
obtained from the Chandra Data Archive. This work made
use of data supplied by the STFC funded UK Swift Science
Data Centre at the University of Leicester. This paper made
use of data products from the JCMT Science Archive (JSA)
project. The JSA is a collaboration between the James Clerk
Maxwell Telescope (JCMT) and the Canadian Astronomy
Data Center (CADC). The JCMT is operated by the Joint
Astronomy Centre on behalf of the Science and Technology
Facilities Council of the United Kingdom, the Netherlands
Organisation for Scientific Research, and the National Research
Council of Canada. The CADC is operated by the
National Research Council of Canada with the support of
the Canadian Space Agency.

\end{document}